\documentclass[12pt]{iopart}

\usepackage{iopams}  

\usepackage{graphicx}

\def\be{\begin{equation}}
\def\ee{\end{equation}}
\def\bea{\begin{eqnarray}}
\def\eea{\end{eqnarray}}

\begin{document}

\title{On the Decay of Soliton Excitations}

\author{H Weigel}

\address{Fachbereich Physik, Siegen University, 57068 Siegen, Germany}
\ead{weigel@physik.uni-siegen.de}

\begin{abstract}
In field theory the scattering about spatially extended objects, such
as solitons, is commonly described by small amplitude fluctuations.
Since soliton configurations often break internal symmetries, excitations
exist that arise from quantizing the modes that are introduced to
restore these symmetries. These modes represent collective distortions and
cannot be treated as small amplitude fluctuations. 
Here we present a method to embrace their contribution to the
scattering matrix. In essence this allows us to compute the decay widths
of such collective excitations. As an example we consider the Skyrme model
for baryons and explain that the method helps to solve the long--standing
Yukawa problem in chiral soliton models.
\end{abstract}


\section{Statement of the problem}

Phase shifts, $\delta$, are essential tools to probe external forces 
in field theory because -- among other applications -- they describe the 
response to background potentials. In turn they play an important role for 
determining Casimir (or vacuum polarization) energies {\it cf.\@}
ref.~\cite{Graham:2002fi},
\begin{equation}
E_{\rm vac}\sim \int \frac{dk}{2\pi}\, \omega_k\,
\frac{d}{dk}\,\delta(k) + E_{\rm c.t.}
\label{eq:evac}
\end{equation}
of extended objects, such as solitons $\Phi_{\rm cl}$. Usually these phase shifts
are computed from small amplitude fluctuations about $\Phi_{\rm cl}$. Solitons 
often break symmetries of the fundamental theory. For example, the Skyrme
hedgehog soliton is not rotationally invariant. This applies to both, coordinate
and flavor rotations. In turn this gives rise to large amplitude fluctuations
in these directions. The symmetries are restored by canonically quantizing
collective coordinates that parameterize the orientation of the soliton 
in the corresponding spaces. The so--constructed states correspond to 
resonances of the soliton and their Yukawa exchanges contribute significantly
to the scattering data. Yet, this important contribution is not necessarily 
captured by the small amplitude fluctuations. Here I will exemplify their 
incorporation within the Skyrme model.

There is a serious problem for computing properties of resonances soliton 
models. Commonly the coupling of resonances to mesons is described by a Yukawa 
interaction of the generic structure
\begin{equation}
\Gamma_{\rm int}[\psi_{B^\prime},\psi_B,\phi]
\sim g\, \int d^4x\, \bar{\psi}_{B^\prime}\,\phi\,\psi_B\,,
\label{eq:yukawa}
\end{equation}
where $B^\prime$ is the resonance that might decay into the (ground) state
$B$ and meson $\phi$ and $g$ is a coupling constant. It is crucial that this 
interaction is \emph{linear} in $\phi$!
If $\phi$ is pseudoscalar, this interaction yields the 
decay width (residuum of the scattering amplitude) 
$\Gamma(B^\prime\to B\phi)\propto g^2 |\vec{p}_\phi|^3$,
with $\vec{p}_\phi$ being the momentum of the outgoing meson.
Soliton models, however, are based on action functionals of \emph{only} 
meson degrees of freedom, $\Gamma=\Gamma[\Phi]$. They contain
classical (static) soliton solutions, $\Phi_{\rm sol}$, that are 
identified as baryons whose interaction with the mesons
is described by the (small) meson fluctuations about the soliton: 
$\Phi=\Phi_{\rm sol}+\phi$. By pure definition of the stationary condition, 
the expansion of $\Gamma[\Phi]$ about $\Phi_{\rm sol}$ does not have 
a term that is linear in $\phi$ to be interpreted as Yukawa
interaction, eq.~(\ref{eq:yukawa}). This puzzle has become famous as the 
Yukawa problem in soliton models. Hence the resonance properties must 
be extracted from meson baryon scattering amplitudes. In 
soliton models two--meson processes acquire contributions from the second 
order term 
\begin{equation}
\Gamma^{(2)}=\frac{1}{2}\,\phi \cdot\,
\frac{\delta^2 \Gamma[\Phi]}{\delta^2 \Phi}\Big|_{\Phi=\Phi_{\rm sol}}\,
\hspace{-0.3cm}\cdot\phi\quad .
\label{eq:secorder}
\end{equation}
This also represents an expansion in $N$, the number of internal degrees 
of freedom (color in strong interactions): $\Gamma=\mathcal{O}(N)$ and 
$\Gamma^{(2)}=\mathcal{O}(N^0)$. Terms $\mathcal{O}(\phi^3)$ vanish 
when $N\to\infty$. Thus $\Gamma^{(2)}$ contains all 
large--$N$ information about the contribution of resonances
to scattering data. 

The large--$N$ expansion is systematic but a low order truncation is not 
necessarily reliable at the physical point and it is very challenging to 
reliably compute subleading contributions. Presumably resonance exchanges 
contribute significantly in that regime. To probe the reliability of the 
computed resonance contributions we transform the above statement into a 
consistency condition: For $N\to\infty$ any valid computation of hadronic 
decay widths in soliton models \emph{must} identically match the result 
obtained from $\Gamma^{(2)}$.  Unfortunately, the most prominent baryon resonance, 
the $\Delta$ isobar, becomes degenerate with the nucleon as $N\to\infty$. It is 
stable in that limit and its decay is not subject to the just described 
litmus--test. The situation is more interesting in soliton models for flavor 
$SU(3)$. In the so--called rigid rotator approach (RRA), that generates baryon 
states as (flavor) rotational excitations of the soliton, exotic resonances emerge 
that dwell in the anti--decuplet representation of flavor $SU(3)$\cite{Theta}. 
The most discussed (and disputed) such state is the $\Theta^+$ pentaquark with 
zero isospin and strangeness $S=+1$. When $N\to\infty$ the mass difference
between anti--decuplet states and the nucleon does not vanish. So the properties 
of $\Theta^+$ predicted from any model treatment must also be seen in the 
quantizing of the strangeness degrees of freedom based on the harmonic 
approximation, $\Gamma^{(2)}$. This (seemingly alternative) quantization is 
called the bound state approach (BSA) for reasons that will become obvious later.  
The above discussed litmus--test requires that the BSA and RRA give identical 
results for $\Theta^+$ as $N\to\infty$. This did not seem to be true 
and it was argued that the prediction of pentaquarks would be a mere artifact of the 
RRA~\cite{It04}. We will show that this conclusion is premature. Furthermore the 
comparison between the BSA and RRA provides an unambiguous computation of pentaquark 
widths: It differs substantially from approaches~\cite{Di97} that adopted 
transition operators for $\Theta^+\to KN$ from the axial current.

This presentation is based on ref.~\cite{Wa05} which should be 
consulted for further details.

\section{The model}

For simplicity we consider the Skyrme model~\cite{Sk61} as a 
particular example for chiral soliton models. However, we stress that our 
qualitative results generalize to \emph{all} chiral soliton models 
because these results solely reflect the treatment of the model 
degrees of freedom. 

Chiral soliton models are functionals of the chiral field, $U$, the 
non--linear realization of the pseudoscalar mesons\footnote{Repeated 
indices are summed as:
$a,b,c,\ldots = 1,\ldots,8$, $\alpha,\beta,\gamma,\ldots = 4,\ldots,7$ 
and $i,j,k,\ldots = 1,2,3$.}, $\phi_a$
\be
U(\vec{x\,},t)={\rm exp}\left[\frac{i}{f_\pi}
\phi_a(\vec{x\,},t)\lambda_a\right]\,,
\label{chiralfield}
\ee
with $\lambda_a$ being the Gell--Mann matrices of $SU(3)$. 
We split the action into three pieces
$ \Gamma = \Gamma_{SK} + \Gamma_{WZ} + \Gamma_{SB}$.
The first term represents the Skyrme model action
\be
\Gamma_{SK} =
\int d^4 x\, {\rm tr}\,\left\{ \frac{f^2_\pi}{4} 
\left[ \partial_\mu U \partial^\mu U^\dagger \right]
+ \frac{1}{32\epsilon^2} \left[ [U^\dagger \partial_\mu U , 
U^\dagger \partial_\nu U]^2\right] \right\} \, .
\label{Skmodel}
\ee
Here $f_\pi=93{\rm MeV}$ is the pion decay constant and $\epsilon$ is 
the Skyrme parameter. The two--flavor version of the Skyrme 
model suggests $\epsilon=4.25$ to reproduce the $\Delta$--nucleon 
mass difference\footnote{To ensure that the (perturbative) $n$--point 
functions scale as $N^{1-n/2}$~\cite{tH74} we substitute 
$f_\pi=93{\rm MeV}\sqrt{N/3}$ and $\epsilon=4.25\sqrt{3/N}$ in the study 
of the $N$ dependence.}. The QCD anomaly is incorporated via the 
Wess--Zumino action~\cite{Wi83} 
\be
\Gamma_{WZ} = - \frac{i N}{240 \pi^2}
\int d^5x \ \epsilon^{\mu\nu\rho\sigma\tau}
{\rm tr}\,\left[\alpha_\mu\alpha_\nu\alpha_\rho\alpha_\sigma\alpha_\tau\right] \,,
\label{WZ}
\ee
with $\alpha_\mu = U^\dagger \partial_\mu U$. 
The flavor symmetry breaking terms are contained in $\Gamma_{\rm SB}$ 
\be
\hspace{-2cm}
\Gamma_{SB} = \frac{f_\pi^2}{4}\, \int d^4x\,
{\rm tr}\, \left[{\cal M}\left(U + U^\dagger - 2 \right)\right]
\,,\quad
\mathcal{M}=\mbox{diag}\left(m_\pi^2,m_\pi^2 ,2m_K^2-m_\pi^2\right)\,.
\label{SB}
\ee
We do not include terms that distinguish between pion and kaon decay 
constants even though they differ by about 20\% empirically. This omission 
is a matter of convenience and leads to an underestimation 
of symmetry breaking effects~\cite{We90} which approximately can be 
accounted for by rescaling the kaon mass $m_K\to m_Kf_K/f_\pi$. 
The action has a topologically non--trivial
classical solution, the famous hedgehog soliton
\be
\Phi_{\rm sol}\,\sim\,U_0(\vec{x\,})
={\rm exp}\left[i\vec{\lambda\,}\cdot\hat{x} F(r)\right]\,,
\quad r=|\vec{x\,}|
\label{hedgehog}
\ee
embedded in the isospin subspace of flavor $SU(3)$. The chiral angle, $F(r)$ 
solves the classical equation of motion subject to the boundary condition
$F(0)-F(\infty)=\pi$ ensuring unit winding (baryon) number. 
In the RRA baryon states are generated by canonically quantizing collective
coordinates $A\in SU(3)$ that describe the (spin) flavor orientation
of the soliton, $A(t)U_0(\vec{x})A^\dagger(t)$. The resultant eigenstates
may be classified according to $SU(3)$ multiplets; see ref.~\cite{We96}
for a review.

\section{Small amplitude fluctuations in the $P$--wave channel with strangeness}

As motivated in chapter~1, we introduce fluctuations 
$\phi\,\sim\,\eta_\alpha(\vec{x\,},t)$
\be
U(\vec{x\,},t)=  \sqrt{U_0(\vec{x\,})}\,
{\rm exp}\left[\frac{i}{f_\pi}\lambda_\alpha\eta_\alpha(\vec{x\,},t)\right]
\sqrt{U_0(\vec{x\,})}\,,
\label{ckfluct}
\ee
for the kaon fields~\cite{Ca85}. Expanding the action in powers 
of these fluctuations yields $\Gamma^{(2)}$ at first non--zero order.
The $P$--wave mode is characterized by a single radial function
\be
\pmatrix{
\eta_4+i\eta_5\cr
\eta_6+i\eta_7}_{P}(\vec{x\,},t)\,
=\int_{-\infty}^\infty\, d\omega\, {\rm e}^{i\omega t}\,
\eta(r,\omega)\,\hat{x}\cdot\vec{\tau}\,\chi(\omega)\,.
\label{pwave}
\ee
In future we will omit the argument for the Fourier 
frequency. Upon quantization the components of the two--component 
iso--spinor $\chi$ are elevated to creation-- and annihilation 
operators. It is straightforward to deduce the Schr\"odinger type 
equation 
\be
\hspace{-2cm}
h^2\,\eta(r)+\omega\left[2\lambda(r)-\omega M_K(r)\right]\eta(r)=0
\quad {\rm with} \quad 
h^2=-\frac{d^2}{dr^2}-\frac{2}{r}\frac{d}{dr}+V_{\rm eff}(r)\,.
\label{scndorder}
\ee
The radial functions arise from the chiral angle $F(r)$ and may
be taken from the literature~\cite{Ca85}. 
The equation of motion~(\ref{scndorder}) is not invariant under particle
conjugation $\omega\leftrightarrow-\omega$, and thus different for
kaons ($\omega>0$) and anti--kaons ($\omega<0$). This difference 
is caused by $\lambda(r)\ne0$ which originates from $\Gamma_{WZ}$.  
Equation~(\ref{scndorder}) has a bound state solution (hence the 
notion \emph{bound state approach}) at $\omega=-\omega_\Lambda$ which
equals the mass difference between the 
$\Lambda$--hyperon and the nucleon in the large--$N$ limit. As this 
energy eigenvalue is negative it corresponds to a kaon, {\it i.e.\@} it 
carries strangeness $S=-1$. In the symmetric case ($m_K=m_\pi$) the bound 
state is the zero mode of $SU(3)$ flavor symmetry. Since $\Gamma_{WZ}$
moves the potential bound state with $S=+1$ to $\omega_\Theta>m_K$ we
expect a resonance structure in that channel. The corresponding phase shift
is shown in the left panel of figure~\ref{fig1}.
No clear resonance structure is visible; the phase shifts
hardly reach $\pi/2$. The absence of such a
resonance has previously lead to the premature criticism that
there would not exist a bound pentaquark in the
large--$N$ limit~\cite{It04}.

\section{Constraint fluctuations in the flavor symmetric case}

We couple the fluctuations to the collective
excitations by generalizing eq.~(\ref{ckfluct}) to
\be
U(\vec{x\,},t)= A(t)\sqrt{U_0(\vec{x\,})}\,
{\rm exp}\left[\frac{i}{f_\pi}\lambda_\alpha
\widetilde{\eta}_\alpha(\vec{x\,},t)\right]
\sqrt{U_0(\vec{x\,})}A^\dagger(t)\,.
\label{ckfluctconst}
\ee
These fluctuations dwell in the intrinsic system as they rotate along with the 
soliton. The kaon $P$--wave is subject to the modified integro--differential 
equation
\bea
&&\hspace{-2.3cm}
h^2\widetilde{\eta}(r)+
\omega\left[2\lambda(r)-\omega M_K(r)\right]\widetilde{\eta}(r)=
-z(r)\left[\int_0^\infty r^{\prime2}dr^\prime z(r^\prime)
2\lambda(r^\prime)\widetilde{\eta}(r^\prime)\right]\cr
&&\hspace{-1.6cm}\times
\left[2\lambda(r)-\left(\omega+\omega_0\right)M_K(r)
-\omega_0\left(\frac{X_\Theta^2}{\omega_\Theta-\omega}
+\frac{X_\Lambda^2}{\omega}\right)
\left(2\lambda(r)-\omega_0M_K(r)\right)\right]\,,\hspace{0.5cm}
\label{Yukawa1}
\eea
for the flavor symmetric case\footnote{The more complicated case 
$m_K\ne m_\pi$ is at length discussed in ref.~\cite{Wa05}.}. The radial 
function $\widetilde{\eta}(r)$ is defined according to eq.~(\ref{pwave}) and 
$z(r)=\sqrt{4\pi}\frac{f_\pi}{\sqrt{\Theta_K}}\, {\rm sin}\frac{F(r)}{2}$ is
the collective mode wave--function normalized with respect to the moment of 
inertia for flavor rotations into strangeness direction, 
$\Theta_K=f_\pi^2 \int d^3r M_K(r)\,{\rm sin}^2\frac{F(r)}{2}=\mathcal{O}(N)$.
The non--local terms reflect the constraint 
$\int dr r^2 z(r)M_K(r)\widetilde{\eta}(r)=0$ which avoids double counting 
of rotational modes in strangeness direction. The interesting coupling is
\be
\hspace{-1.5cm}
H_{\rm int}=\frac{2}{\sqrt{4\pi\Theta_K}}\,
d_{i\alpha\beta}\, D_{\gamma\alpha}R_\beta\,
\int d^3r\, z(r)\left[2\lambda(r)-\omega_0 M_K(r)\right]
\hat{x}_i \widetilde{\xi}_\gamma(\vec{x\,},t)\,,
\label{hint}
\ee
where $\widetilde{\xi}_a=D_{ab}\widetilde{\eta}_b$ are the fluctuations in the 
laboratory frame, that we actually detect in $KN$ scattering. The collective 
coordinates are parameterized via the adjoint representation 
$D_{ab}(A)=\frac{1}{2}\,{\rm tr}\,\left[\lambda_a A \lambda_b A^\dagger\right]$
and the $SU(3)$ generators $R_a$. Integrating out the collective degrees of 
freedom induces the separable potential
\be
\frac{\left|\langle\Theta|H_{\rm int}|(KN)_{I=0}\rangle\right|^2}
{\omega_\Theta-\omega}
+\frac{\left|\langle\Lambda|H_{\rm int}|(KN)_{I=0}\rangle\right|^2}
{\omega_\Lambda+\omega}\,.
\label{potential}
\ee
These matrix elements concern the $T$--matrix elements in the
laboratory frame. For the $\Theta^+$ channel it is identical to the
one in the intrinsic system~\cite{Ha84}. Thus we may directly add the 
exchange potential, eq.~(\ref{potential}) to the Hamiltonian for the 
intrinsic fluctuations. We define matrix elements of collective 
coordinate operators
\be
\hspace{-1.5cm}
\langle \Theta^+| d_{3\alpha\beta} D_{+\alpha}R_\beta|n\rangle 
=:X_\Theta\sqrt{\frac{N}{32}} \quad {\rm and}\quad
\langle \Lambda| d_{3\alpha\beta}
D_{-\alpha}R_\beta|p\rangle=:X_\Lambda\sqrt{\frac{N}{32}}\,,
\label{ddr}
\ee
to end up with eq.~(\ref{Yukawa1}). The first factor in the coefficient
$\omega_0=2\left(\frac{2}{\sqrt{\Theta_K}}\sqrt{\frac{N}{32}}\right)^2=
\frac{N}{4\Theta_K}$ arises in the equation of motion because the potential, 
eq.~(\ref{potential}) is quadratic in the fluctuations. The remaining 
(squared) factors stem from the definitions of $X_{\Theta,\Lambda}$ and the 
constant of proportionality in $H_{\rm int}$. The $X_{\Theta,\Lambda}$
must be computed with the methods provided in ref.~\cite{Ya88} but generalized 
to arbitrary (odd) $N$~\cite{Wa05}. For $N\to\infty$ we have
$X_\Theta\to1$ and $X_\Lambda\to0$. From the orthogonality conditions of the 
equation of motion~(\ref{scndorder}) we straightforwardly verify that
in this limit
\be
\hspace{-1.0cm}
\widetilde{\eta}(r)=\eta(r)-az(r)
\qquad {\rm with} \qquad
a=\int_0^\infty dr r^2\, z(r) M_K(r)\eta(r)\,.
\label{solution}
\ee
solves eq.~(\ref{Yukawa1}). This is essential because, 
as $z(r)$ is localized in space, $\eta$ and $\widetilde{\eta}$ have 
identical phase shifts! Hence the large--$N$ consistency condition discussed 
in the introduction is indeed satisfied. The physics becomes more 
transparent when considering the background wave--function $\overline{\eta}(r)$ 
that solves eq.~(\ref{Yukawa1}) for $X_\Theta\equiv X_\Lambda\equiv0$. 
{\it i.\@ e.\@} the collective excitations are decoupled. We stress that
$\overline{\eta}(r)$ is a purely large--$N$ quantity that may be
obtained by demanding the BSA wave--function to be orthogonal to the
collective mode. In doing so, the asymptotic behavior of $\overline{\eta}(r)$ 
gives the background phase shift shown in figure~\ref{fig1}. Subsequently we 
may again switch on the exchange contributions, eq.~(\ref{potential}). The 
additional separable potential augments the phase shift by
\be
{\rm tan}\left(\delta_{\rm R}(\omega)\right)=
\frac{\Gamma(\omega_k)/2}{\omega_\Theta-\omega+\Delta(\omega)}\,.
\label{resformula}
\ee
Here $\omega_\Theta=\frac{N+3}{4\Theta_K}$ is the RRA result for the 
excitation energy of $\Theta$.  This phase shift exhibits the canonical 
resonance structure with the width and the pole shift
\bea
\Gamma(\omega)&=&2k\omega_0 X_\Theta^2
\left|\int_0^\infty r^2dr\, z(r)2\lambda(r)
\overline{\eta}(r,\omega)\right|^2\,,
\label{width1} \\
\Delta(\omega)&=&\frac{1}{2\pi\omega}\,{\cal P}\,
\int_0^\infty q dq\,\left[
\frac{\Gamma(\omega_q)}{\omega-\omega_q}
+\frac{\Gamma(-\omega_q)}{\omega+\omega_q}\right]\,\,,
\quad \big(\omega_q=\sqrt{q^2+m_K^2}\big) \,.
\label{poleshift}
\eea
We have numerically verified that in the large--$N$ limit 
with $X_\Theta^2=1$, the phase shift from eq.~(\ref{resformula}) is 
identical to what is labeled resonance phase shift in figure~\ref{fig1}, 
that we calculated as the difference between the total ($\eta$) and 
background ($\overline{\eta}$) phase shifts. It stems from
a completely different computation: While all phase 
shifts shown in figure~\ref{fig1} result solely from large--$N$ computations,
eq.~(\ref{resformula}) yields an $N$ dependent phase shift that for
$N\to\infty$ reproduces the resonance phase shift. This clearly 
shows that contrary to earlier criticisms~\cite{It04} the large $N$ 
pentaquark channel indeed resonates! It furthermore suggests that the 
small amplitude approach~(\ref{eq:secorder}) might give insufficient results 
for scattering data.

\section{Results}

\begin{figure}[t]
\centerline{
\includegraphics[width=4.0cm,height=6.2cm,angle=270]{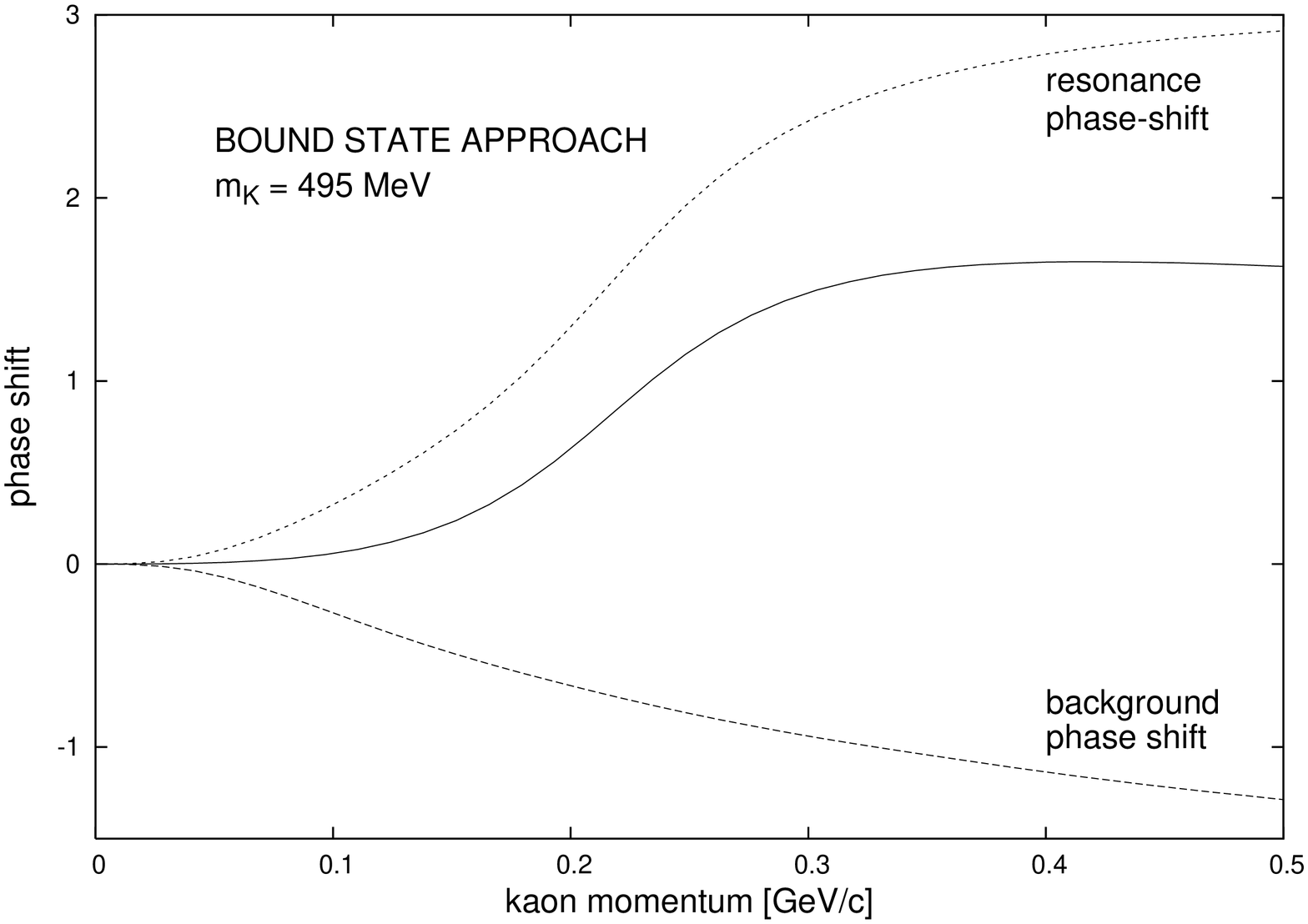}
\hspace{1cm}
\includegraphics[width=4.0cm,height=6.2cm,angle=270]{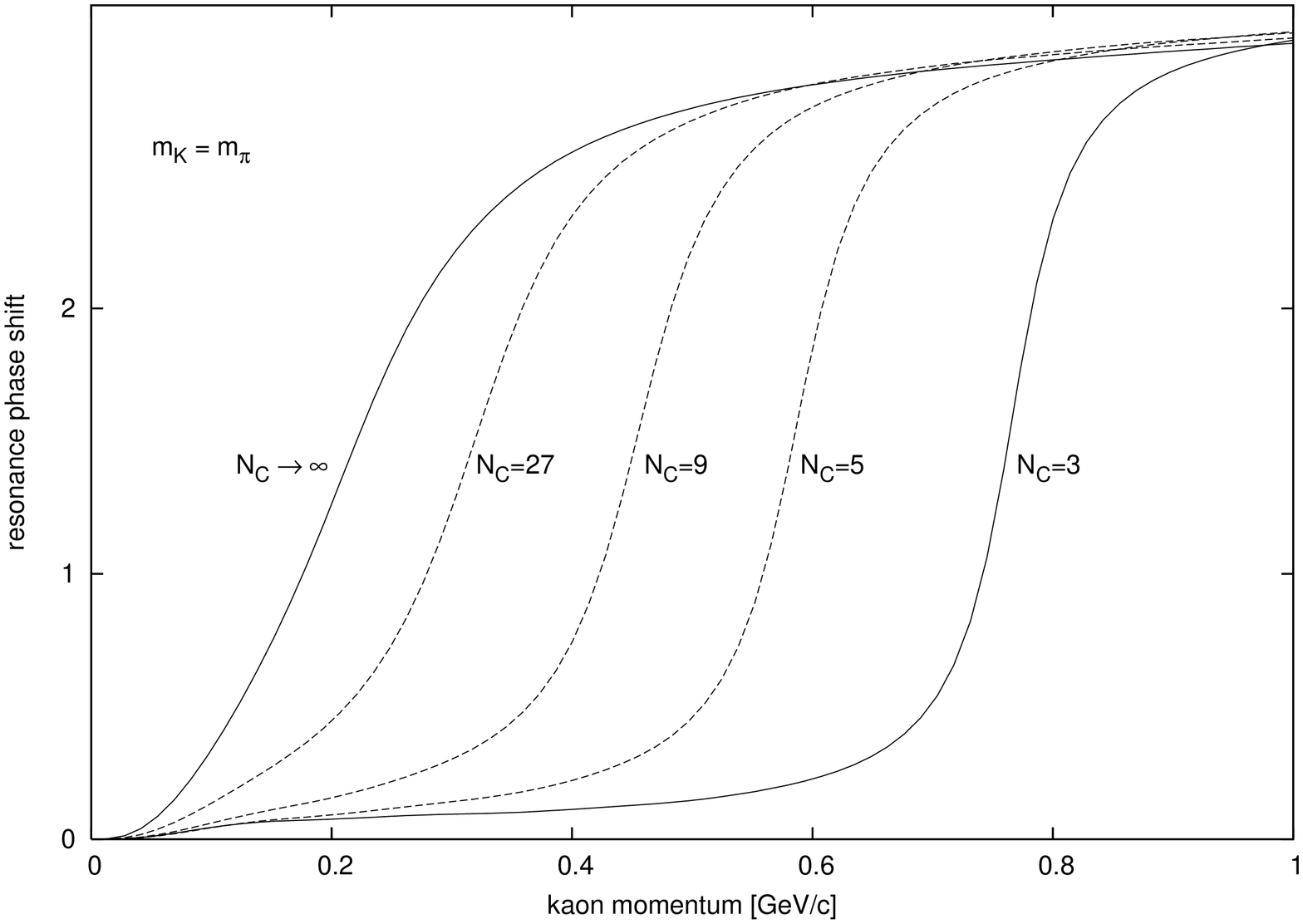}}
\leftline{~\hspace{-1.7cm}\begin{minipage}[l]{16.4cm}
\caption{\label{fig1}Left panel:
Large $N$ $P$--wave phase shifts with strangeness $S=+1$. The full line shows 
the phase extracted from eq.~(\ref{scndorder}). Background and resonance 
phase shifts are defined in the text.
Right panel: The resonance phase shift for various $N$ and $m_K=m_\pi$.}
\end{minipage}}
\end{figure}

In figure~\ref{fig1} we show the resonance phase shift computed from
eq.~(\ref{resformula}) for various values of $N$. First we observe that the 
resonance position quickly moves towards larger energies as $N$ decreases. This 
is mainly due to the strong $N$ dependence of $\omega_\Theta$: For $N=3$ it
is twice as large as in the limit $N\to\infty$. The pole shift $\Delta$
is quite small (some ten MeV) so $\omega_\Theta$ is indeed
a reliable estimate of the resonance energy. Second, the resonance
becomes shaper as $N$ decreases. 
To discuss the quantitative results we now include
flavor symmetry breaking effects. Then the resonance position changes to
\be
\omega_\Theta=\frac{1}{2}\left[\sqrt{\omega_0^2+\frac{3\Gamma}{2\Theta_K}}
+\omega_0\right]+\mathcal{O}\left(\frac{1}{N}\right)\,.
\label{omt}
\ee
where $\Gamma=\mathcal{O}(N)$ is a functional of the soliton that is 
proportional to the meson mass difference, $m_K^2-m_\pi^2$. The 
$\mathcal{O}\left(1/N\right)$ piece is sizable for $N=3$ and we 
compute it in the scenario of ref.\cite{Ya88}. We then find 
$\omega_\Theta\approx700{\rm MeV}$; taking model dependencies into
account we expect the pentaquark to be about $600\ldots900{\rm MeV}$ 
heavier than the nucleon.

For the width calculations there are two principle differences to 
eq.~(\ref{width1}). First, the interaction Hamiltonian acquires an additional term 
\be
H_{\rm int}^{\rm sb}=
\left(m_K^2-m_\pi^2\right)d_{i\alpha\beta}D_{\gamma\alpha}D_{8\beta}
\int d^3r\,z(r)\gamma(r)\widetilde{\xi}_{\gamma}(\vec{x\,},t)\hat{x}_i\,,
\label{hintsb}
\ee
The radial function $\gamma(r)$ is again given in terms of the chiral 
angle~\cite{Wa05}. Second, the $X_\Lambda$ does not vanish as $N\to\infty$
and the $R$--matrix formalism is always two dimensional. Nevertheless, the 
large--$N$ solution is always of the form~(\ref{solution}) and the BSA
phase shift is recovered. The width function turns to
\be
\hspace{-1.0cm}
\Gamma(\omega)=2k\omega_0 \left|\int_0^\infty \hspace{-1mm} r^2dr\, z(r)
\left[2X_\Theta\lambda(r) +\frac{Y_\Theta}{\omega_0}\left(m_K^2-m_\pi^2\right)
\right]\overline{\eta}(r,\omega)\right|^2,
\label{widthsb}
\ee
where $X_\Theta$ and $Y_\Theta=\sqrt{8N/3} 
\langle\Theta^+|d_{3\alpha\beta}D_{+\alpha}D_{8\beta}|n\rangle$ are to be 
computed in the RRA approach with full inclusion of flavor symmetry breaking
effects.

\begin{figure}[t]
\centerline{
\includegraphics[width=4.0cm,height=6.2cm,angle=270]{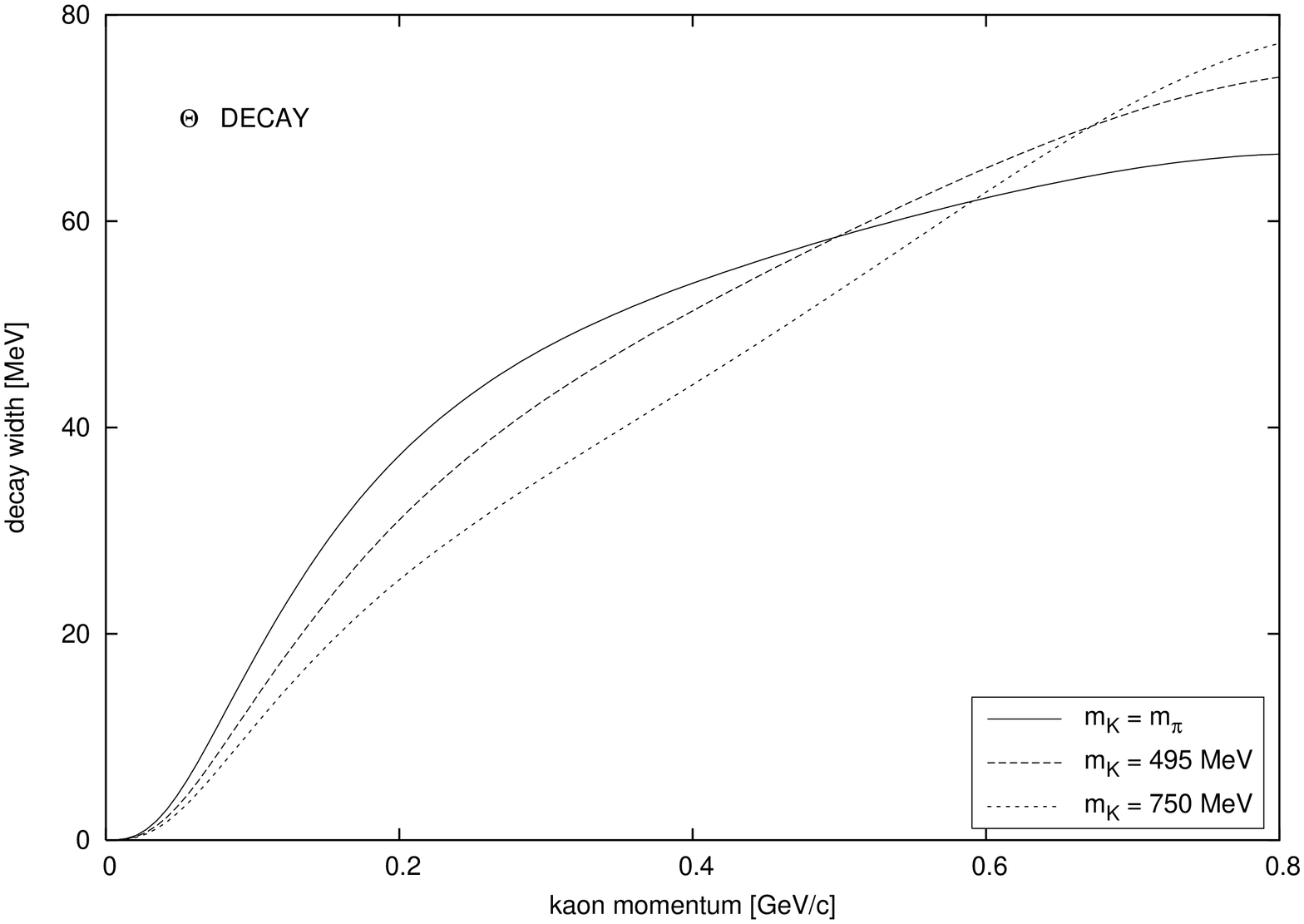}\hspace{1cm}
\includegraphics[width=4.0cm,height=6.2cm,angle=270]{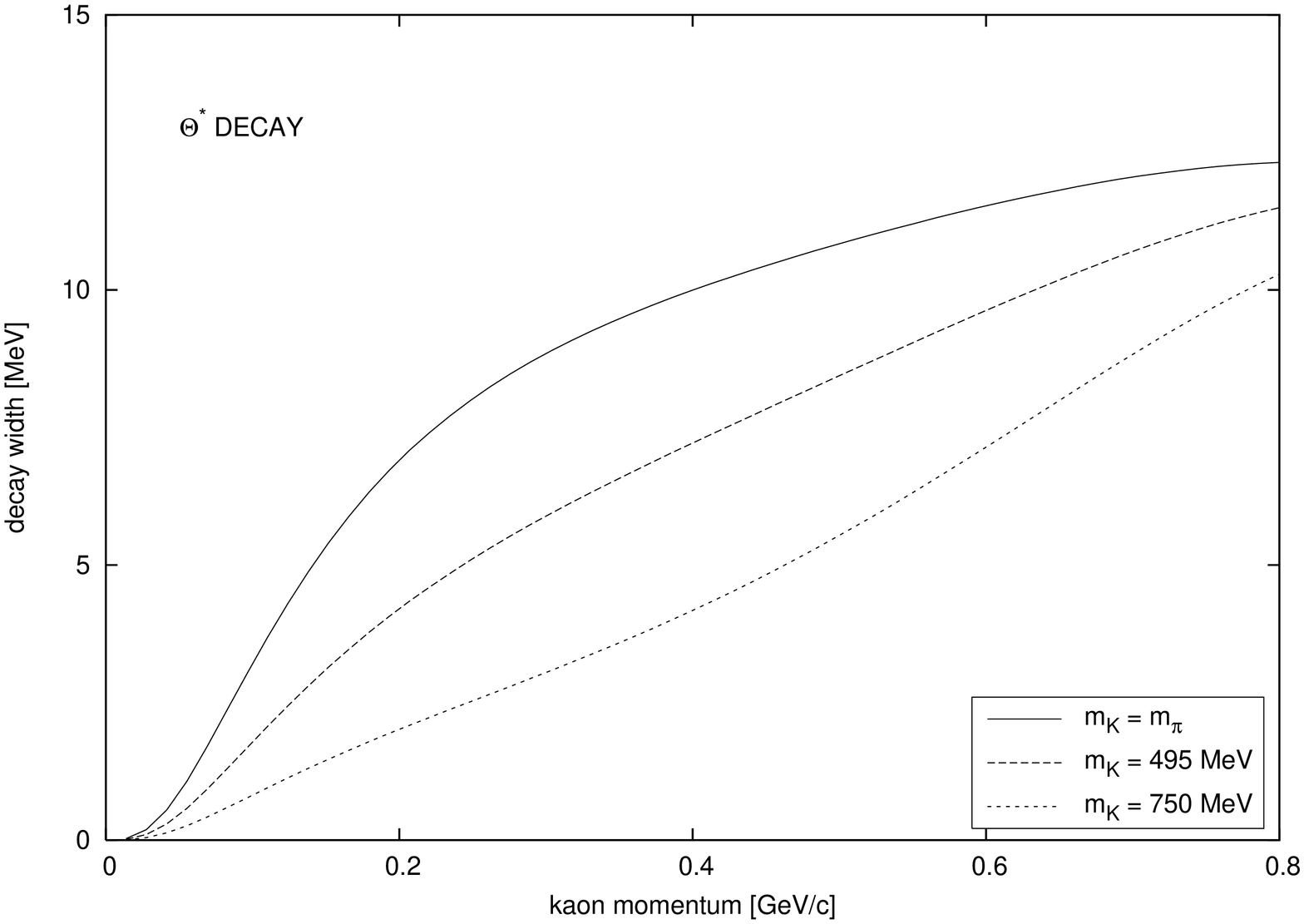}}
\leftline{~\hspace{-1.5cm}\begin{minipage}[l]{16.0cm}
\caption{\label{fig3}Model prediction for the
width, $\Gamma(\omega)$ of $\Theta^+$ (left) and $\Theta^{*+}$ (right)
for $N=3$ for three values of the kaon mass.
Note the unequal scales.}
\end{minipage}}
\end{figure}

This width function is shown (for $N=3$) in figure~\ref{fig3} for 
$\Theta$ and its isovector partner $\Theta^*$. The latter merely 
requires the appropriate modification of the matrix elements in 
eq.~(\ref{ddr}). The $k^3$ behavior of the width
function, as suggested by the model, eq.~(\ref{eq:yukawa}) is 
reproduced only right above threshold, afterwards it levels off.
Somewhat surprising, the width of the non--ground 
state pentaquark is smaller than that of the lowest lying 
pentaquark. Our particular model yields
$\Gamma_\Theta\approx{40}{\rm MeV}$ and 
$\Gamma_{\Theta^*}\approx{20}{\rm MeV}$. We note that there
are certainly model ambiguities in these results.

\section{Conclusions}

To exemplify the role resonance exchanges play for the computation of
scattering data in soliton models we have discussed $KN$ scattering 
in the $S=+1$ channel which contains the potential~\cite{Hi07} $\Theta^+$ 
pentaquark, a state predicted as a flavor rotational excitation.
Though the approach via small amplitude fluctuations suggests otherwise, 
the $\Theta$ emerges as a genuine resonance. A central result is the width 
function for $\Theta\to KN$. In the flavor symmetric case it contains only 
a \emph{single} collective coordinate operator and is thus very different 
from estimates that extract an effective Yukawa coupling from the axial 
current matrix element~\cite{Di97}. Since our approach matches the 
exact large $N$ result, we must conclude that those axial current scenarios 
are erroneous~\cite{We07} and that soliton models unlikely predict very
narrow pentaquarks.

\section*{Acknowledgments}

This presentation is based on a collaboration with H.\@ Walliser,
whose contribution is highly appreciated. I am very grateful to 
the organizers, in particular M.\@ Bordag for making this an
enjoyable workshop.

\end{document}